\documentclass[3p,times]{elsarticle}

\usepackage[english]{babel} 

\usepackage[%
	pdftitle={Heavy quarks at RHIC and LHC within a partonic transport model},%
	pdfauthor={Jan Uphoff, Oliver Fochler, Zhe Xu, and Carsten Greiner},%
	pdfsubject={Heavy quarks at RHIC and LHC within a partonic transport model},
	pdfstartview=FitH,
	pdfpagemode=UseNone,
	bookmarksopen=true
	]{hyperref}

\usepackage[
   centertags, 
   sumlimits,  
   intlimits,  
   namelimits, 
]{amsmath} %
\usepackage{amssymb}

\begin{document}

\begin{frontmatter}




\title{Heavy quarks at RHIC and LHC within a partonic transport model}


\author[label1]{Jan Uphoff}
\author[label1]{Oliver Fochler}
\author[label2,label1]{Zhe Xu}
\author[label1]{Carsten Greiner}

\address[label1]{Institut f\"ur Theoretische Physik, Johann Wolfgang 
Goethe-Universit\"at Frankfurt, Max-von-Laue-Str. 1, 
D-60438 Frankfurt am Main, Germany}
\address[label2]{Frankfurt Institute for Advanced Studies, Ruth-Moufang-Str. 1, D-60438 Frankfurt am Main, Germany}

\begin{abstract}
Production and space-time evolution of heavy quarks in central and non-central heavy-ion collisions at RHIC and LHC are studied with the partonic transport model \emph{Boltzmann Approach of MultiParton Scatterings} (BAMPS). In addition to the initially created heavy quarks in hard parton scatterings during nucleon-nucleon collisions,  secondary heavy quark production in the quark-gluon plasma is investigated and the sensitivity on various parameters is estimated. In BAMPS heavy quarks scatter with particles of the medium via elastic collisions, whose cross section is calculated with the running coupling and a more precise implementation of Debye screening. In this framework, we compute the elliptic flow and nuclear modification factor of heavy quarks and compare it to the experimental data.
\end{abstract}

\begin{keyword}
Quark-gluon plasma \sep heavy quarks \sep Boltzmann equation \sep elliptic flow \sep nuclear modification factor


\end{keyword}

\end{frontmatter}

\section{Introduction}
Several experimental observations  indicate that a medium of free quarks and gluons -- the quark-gluon plasma (QGP) -- is produced in ultra-relativistic heavy-ion collisions \cite{Adams:2005dq,Adcox:2004mh}. Charm and bottom quarks are an ideal probe for the early stage of these collisions since they can only be created in initial hard parton scatterings of nucleon-nucleon interactions or in the beginning of the QGP phase, where the energy density is still large. After their production they interact with other particles of the medium and can, therefore, reveal important information about the properties of the QGP. Flavor conservation renders them as an unique probe since they are tagged by their flavor even after hadronization.

The experimental data of the elliptic flow $v_2$  and nuclear modification factor $R_{AA}$  of heavy quarks \cite{Abelev:2006db,Adare:2006nq,Adare:2010de} show that the energy loss of charm and bottom quarks is comparable to that of light quarks. Whether this large energy loss is due to collisional or radiative interactions -- or both (or even other effects) -- is under investigation 
(see \cite{Adare:2010de} for a recent overview and comparison with data).

After the introduction of the parton cascade BAMPS we will discuss the production of heavy quarks at RHIC and LHC. In Sec.~\ref{sec:energy_loss} our results on the elliptic flow  and nuclear modification factor at RHIC are discussed and compared to the experimental data.

\section{Parton cascade BAMPS}
\label{sec:bamps}
For the simulation of the QGP we use the partonic transport model \emph{Boltzmann Approach of MultiParton Scatterings} (BAMPS) \cite{Xu:2004mz,Xu:2007aa}, which describes the full space-time evolution of the QGP by solving the Boltzmann equation,
\begin{equation}
\label{boltzmann}
\left ( \frac{\partial}{\partial t} + \frac{{\mathbf p}_i}{E_i}
\frac{\partial}{\partial {\mathbf r}} \right )\,
f_i({\mathbf r}, {\mathbf p}_i, t) = {\cal C}_i^{2\rightarrow 2} + {\cal C}_i^{2\leftrightarrow 3}+ \ldots  \ ,
\end{equation}
for on-shell partons and pQCD interactions. 
Details of the model, the implemented processes, and the employed cross sections can be found in \cite{Xu:2004mz,Xu:2007aa,Uphoff:2010sh}.

\section{Heavy quark production at RHIC and LHC}
In heavy-ion collisions charm and bottom quarks are produced in hard parton scatterings of primary nucleon-nucleon collisions or in the QGP. To estimate the initial heavy quark yield, we use PYTHIA \cite{Sjostrand:2006za} and scale from proton-proton collisions to heavy-ion collisions with the number of binary collisions. Secondary heavy quark production in the QGP is simulated with BAMPS. For the initial gluon distributions, the mini-jet model, the color glass condensate model and also PYTHIA are employed.

\begin{figure}
\begin{minipage}[t]{0.49\textwidth}
\centering
\includegraphics[width=1.0\textwidth]{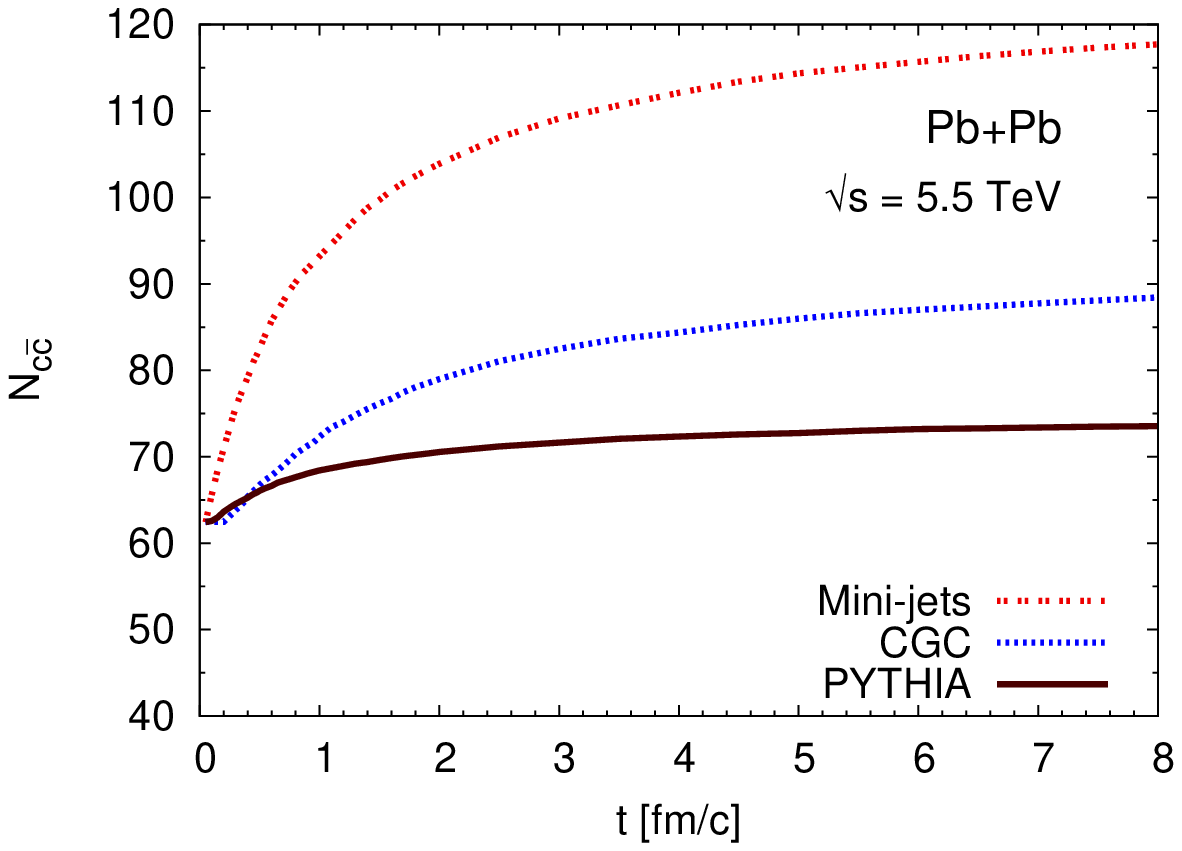}
\end{minipage}
\hfill
\begin{minipage}[t]{0.49\textwidth}
\centering
\includegraphics[width=1.0\textwidth]{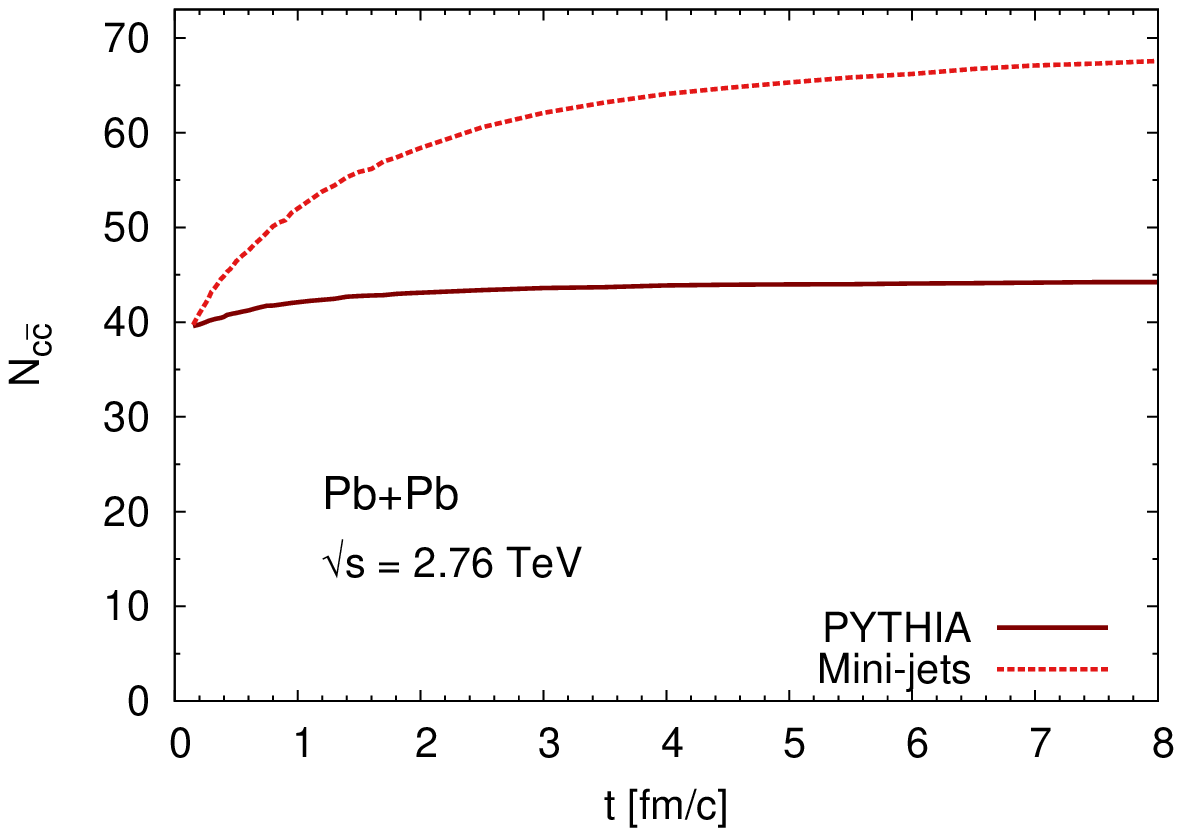}
\end{minipage}
\caption{Number of charm quark pairs produced in a central Pb+Pb collision at LHC with $\sqrt{s_{NN}}=5.5 \, \rm{TeV}$ (left) and $\sqrt{s_{NN}}=2.76 \, \rm{TeV}$ (right) according to BAMPS. The initial parton distributions are obtained with PYTHIA and the mini-jet model (and the color glass condensate for $\sqrt{s_{NN}}=5.5 \, \rm{TeV}$). In all cases the initial charm quarks are sampled with PYTHIA for better comparison.}
\label{fig:charm_prod_276}
\end{figure}

In Au+Au collisions at RHIC with $\sqrt{s_{NN}}=200 \, \rm{GeV}$ between 0.3 and 3.4 charm pairs are produced in the QGP, depending on the model of the initial gluon distribution, the charm mass and whether a $K=2$ factor for higher order corrections of the cross section is employed \cite{Uphoff:2010sh}. This is only a small fraction of the initially produced charm quarks and can be neglected for the most probable scenarios.
At LHC with the much larger initial energy density, secondary charm production is enhanced and not negligible. In Pb+Pb collisions with $\sqrt{s_{NN}}=5.5 \, \rm{TeV}$ between 11 and 55 charm pairs are produced during the evolution of the QGP \cite{Uphoff:2010sh} (see left panel of Fig.~\ref{fig:charm_prod_276}). These values are of the same order as the initial yield. As is shown in the right panel of Fig.~\ref{fig:charm_prod_276}, even in the 2010 run with $\sqrt{s_{NN}}=2.76 \, \rm{TeV}$ between 5 and 28 charm pairs are created during the QGP phase.

Bottom production in the QGP, however is very small both at RHIC and LHC and can be safely neglected. As a consequence, all bottom quarks at these colliders are produced in initial hard parton scatterings.

Further details on heavy quark production can be found in Ref.~\cite{Uphoff:2010sh}.

\section{Elliptic flow and nuclear modification factor of heavy quarks at RHIC}
\label{sec:energy_loss}

The elliptic flow and the nuclear modification factor
\begin{align}
\label{elliptic_flow}
  v_2=\left\langle  \frac{p_x^2 -p_y^2}{p_T^2}\right\rangle \ , \qquad \qquad
R_{AA}=\frac{{\rm d}^{2}N_{AA}/{\rm d}p_{T}{\rm d}y}{N_{\rm bin} \, {\rm d}^{2}N_{pp}/{\rm d}p_{T}{\rm d}y}
\end{align} 
($p_x$ and $p_y$ are the momenta in $x$ and $y$ direction in respect to the reaction plane)
of heavy quarks at mid-rapidity are observables which are experimentally measurable and reflect the coupling of heavy quarks to the medium. A large elliptic flow and a small $R_{AA}$ indicate strong interactions with the medium and a sizeable energy loss. Experimental results reveal that both quantities are of the same order as the respective values for light particles \cite{Abelev:2006db,Adare:2006nq,Adare:2010de}.

The leading order perturbative QCD cross section with a constant coupling $\alpha_s = 0.3$ and the Debye mass for  the $t$ channel screening is too small to build up the large elliptic flow measured at RHIC \cite{Uphoff:2010fz}. 
However, if we take the running of the coupling into account and determine the screening mass from comparison to hard thermal loop calculations, we obtain a $v_2$ and $R_{AA}$, which are much closer to the data.

The following calculations are done analogously to \cite{Gossiaux:2008jv,Peshier:2008bg,Uphoff:2010sy,Bouras:2010yw}. An effective running coupling is obtained from measurements of $e^+e^-$ annihilation and non-strange hadronic decays of $\tau$ leptons \cite{Dokshitzer:1995qm,Gossiaux:2008jv}.
Since the $t$ channel of the $g Q \rightarrow g Q$ cross section is divergent, it is screened with a mass proportional to the Debye mass $m_{D}$:
\begin{align}
\label{t_screening}
   \frac{1}{t} \rightarrow \frac{1}{t-\kappa \, m_{D}^2}
\end{align}
The Debye mass is calculated by the standard definition $m_{D}^2 = 4 \pi \, (1+N_f/6) \, \alpha_s(t) \, T^2$, but with the running coupling for consistency. The prefactor $\kappa$ in Eq.~\ref{t_screening} is mostly set to 1 in the literature without a sophisticated reason. However, one can fix this factor by comparing the energy loss per unit length ${\rm d}E/{\rm d}x$ of the born cross section with $\kappa$ to the energy loss within the hard thermal loop approach to $\kappa \approx 0.2$ \cite{Gossiaux:2008jv,Peshier:2008bg}.

\begin{figure}
\begin{minipage}[t]{0.49\textwidth}
\centering
\includegraphics[width=1.0\textwidth]{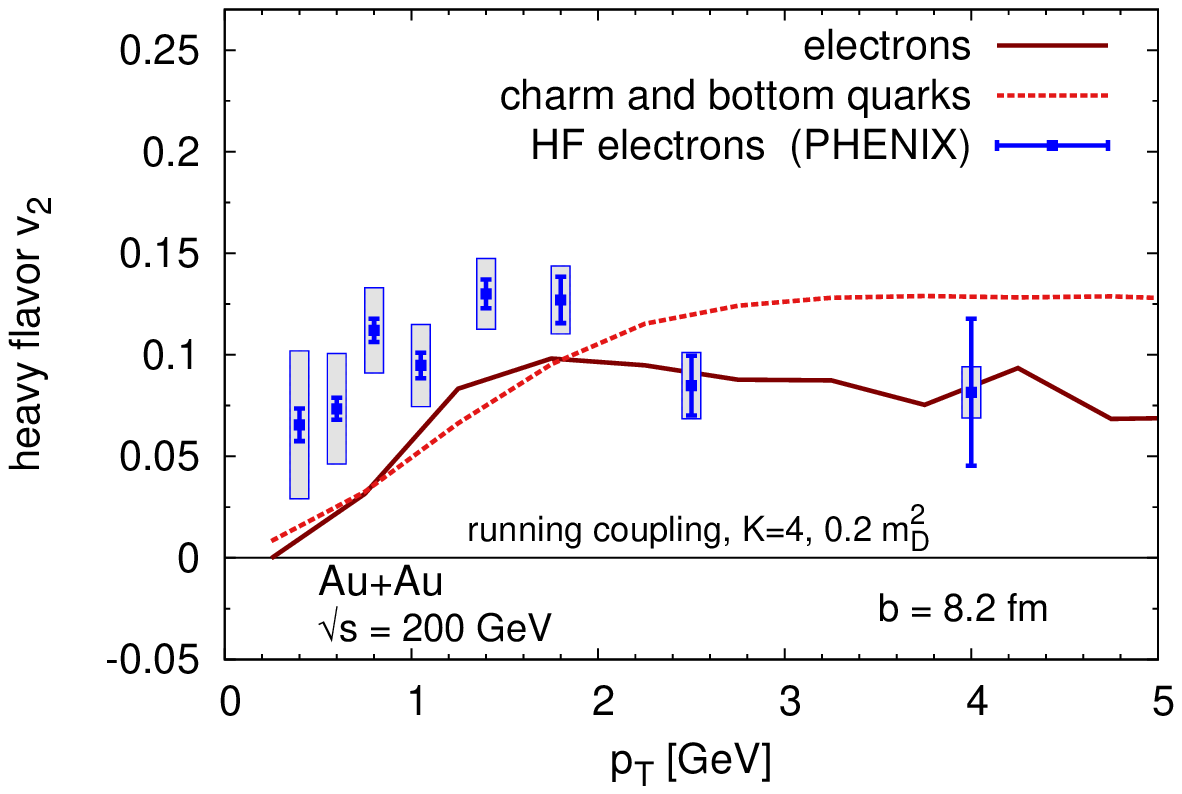}
\end{minipage}
\hfill
\begin{minipage}[t]{0.49\textwidth}
\centering
\includegraphics[width=1.0\textwidth]{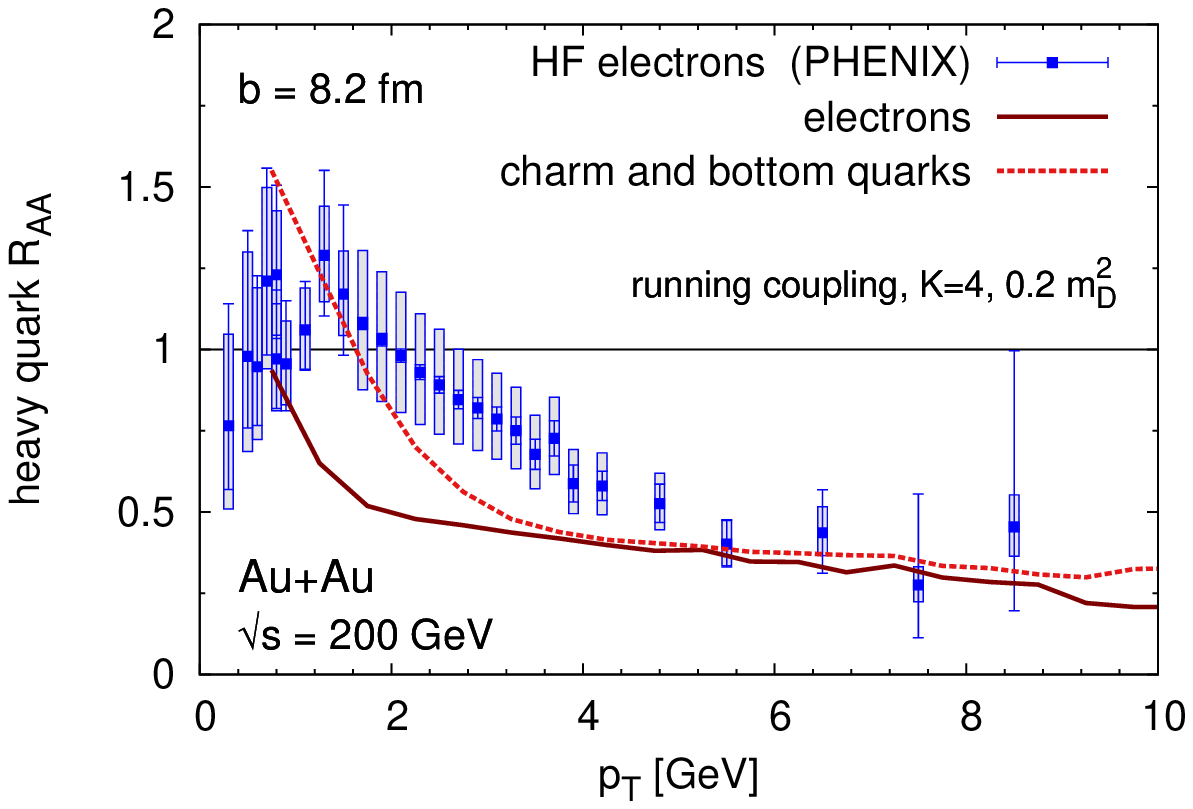}
\end{minipage}
\caption{Elliptic flow $v_2$ (left) and nuclear modification factor $R_{AA}$ (right) of heavy quarks and heavy flavor electrons with pseudo-rapidity $|\eta|<0.35$ at the end of the QGP phase for Au+Au collisions at RHIC with an impact parameter of $b=8.2 \, {\rm fm}$. The cross section of $gQ \rightarrow gQ$ is multiplied with the factor $K=4$. For comparison, data of heavy flavor electrons \cite{Adare:2010de} is shown.}
\label{fig:v2_raa}
\end{figure}

This more accurate treatment increases the cross section of elastic gluon heavy quark scattering by about a factor of 10. Fig.~\ref{fig:v2_raa} shows the elliptic flow $v_2$ and nuclear modification factor $R_{AA}$ for heavy quarks and for heavy flavor electrons as a function of the transverse momentum $p_T$. 
To yield the same values for these variables as the experimental data, the leading order cross section of elastic collisions with the running coupling and improved Debye screening must still be multiplied by a $K$ factor of 4. We assume that this artificial $K$ factor stands for the contribution of radiative energy loss. However, it must be checked if these corrections have indeed a similar effect as a constant $K$ factor of 4. Therefore, the calculation of the next-to-leading order cross section is planned for the near future and will complement $2 \leftrightarrow 3$ interactions for gluons, which are already implemented in BAMPS \cite{Xu:2004mz}.

Especially for high $p_T$ the shape of the $v_2$ curve of heavy quarks is different from the experimental data. The reason for this discrepancy is that, experimentally, due to confinement, not heavy quarks, but heavy flavor electrons are measured. The latter stem from the decay of $D$ and $B$ mesons, which in turn are produced  during hadronization of the QGP by merging of the charm or bottom quark with a light quark. However, despite the hadronization and decay processes heavy flavor electrons still reveal information about heavy quarks. Essentially, the shape of their spectrum is the same as for heavy quarks, but shifted to lower $p_T$ due to the decay process. 

For the description of the hadronization process of charm (bottom) quarks to $D$ ($B$) mesons, we use  Peterson fragmentation \cite{Peterson:1982ak}. The decay to heavy flavor electrons is carried out with PYTHIA. Fig.~\ref{fig:v2_raa} shows that the theoretical curves for heavy flavor electrons are in good agreement with the experimental data for high $p_T$. For low $p_T$, however, Peterson fragmentation is not a good description of the hadronization process and another scheme like coalescence must be employed. In the coalescence picture, the light quarks of the $D$/$B$ mesons contribute also to its elliptic flow or nuclear modification factor, which increases both.
Studies on the $v_2$ and $R_{AA}$ of gluons in BAMPS are presented in \cite{Xu:2008av,Bouras:2008ip,Xu:2010cq,Fochler:2008ts,Fochler:2010wn}.

\section{Conclusions}

In this talk we presented results on heavy quark production as well as on the elliptic flow and nuclear modification factor of heavy quarks.
Charm production in the QGP at RHIC is to a good approximation negligible, but at LHC a significant fraction of the total charm number is produced in the QGP -- even at a center-of-mass energy of $\sqrt{s_{NN}}=2.76 \, \rm{TeV}$. Bottom production in the medium is negligible at RHIC and LHC.
The leading order cross section of heavy quark scatterings with particles from the medium is too small to explain the experimentally measured elliptic flow and nuclear modification factor. However, a more precise implementation of Debye screening and the explicit running of the coupling enhances the cross section and yields results for heavy flavor electrons, which are much closer to the data, although a $K$ factor of 4 must be employed for a good agreement with the data. In the future we will study if this simple multiplication of the cross section with a $K$ factor can indeed account for higher order contributions.

\section*{Acknowledgements}
J.U. would like to thank A. Peshier for stimulating and helpful discussions and the kind hospitality at the University of Cape Town, where part of this work has been done.
The BAMPS simulations were performed at the Center for Scientific Computing of the Goethe University Frankfurt. This work was supported by the Helmholtz International Center for FAIR within the framework of the LOEWE program launched by the State of Hesse.

\bibliographystyle{elsarticle-num}
\bibliography{hq}

\begin{thebibliography}{10}
\expandafter\ifx\csname url\endcsname\relax
  \def\url#1{\texttt{#1}}\fi
\expandafter\ifx\csname urlprefix\endcsname\relax\def\urlprefix{URL }\fi
\expandafter\ifx\csname href\endcsname\relax
  \def\href#1#2{#2} \def\path#1{#1}\fi

\bibitem{Adams:2005dq}
J.~Adams, et~al., Experimental and theoretical challenges in the search for the
  quark gluon plasma: The star collaboration's critical assessment of the
  evidence from rhic collisions, Nucl. Phys. A757 (2005) 102--183.
\newblock \href {http://arxiv.org/abs/nucl-ex/0501009}
  {\path{arXiv:nucl-ex/0501009}}.

\bibitem{Adcox:2004mh}
K.~Adcox, et~al., Formation of dense partonic matter in relativistic nucleus
  nucleus collisions at rhic: Experimental evaluation by the phenix
  collaboration, Nucl. Phys. A757 (2005) 184--283.
\newblock \href {http://arxiv.org/abs/nucl-ex/0410003}
  {\path{arXiv:nucl-ex/0410003}}.

\bibitem{Abelev:2006db}
B.~I. Abelev, et~al., {Transverse momentum and centrality dependence of high-pt
  non-photonic electron suppression in Au+Au collisions at $\sqrt{s_{NN}}$ =
  200 GeV}, Phys. Rev. Lett. 98 (2007) 192301.
\newblock \href {http://arxiv.org/abs/nucl-ex/0607012}
  {\path{arXiv:nucl-ex/0607012}}, \href
  {http://dx.doi.org/10.1103/PhysRevLett.98.192301}
  {\path{doi:10.1103/PhysRevLett.98.192301}}.

\bibitem{Adare:2006nq}
A.~Adare, et~al., {Energy Loss and Flow of Heavy Quarks in Au+Au Collisions at
  $\sqrt{s_{NN}} = 200$~GeV}, Phys. Rev. Lett. 98 (2007) 172301.
\newblock \href {http://arxiv.org/abs/nucl-ex/0611018}
  {\path{arXiv:nucl-ex/0611018}}, \href
  {http://dx.doi.org/10.1103/PhysRevLett.98.172301}
  {\path{doi:10.1103/PhysRevLett.98.172301}}.

\bibitem{Adare:2010de}
A.~Adare, et~al., {Heavy Quark Production in p+p and Energy Loss and Flow of
  Heavy Quarks in Au+Au Collisions at sqrt($s_{NN}$)=200 GeV}\href
  {http://arxiv.org/abs/1005.1627} {\path{arXiv:1005.1627}}.

\bibitem{Xu:2004mz}
Z.~Xu, C.~Greiner, {Thermalization of gluons in ultrarelativistic heavy ion
  collisions by including three-body interactions in a parton cascade}, Phys.
  Rev. C71 (2005) 064901.
\newblock \href {http://arxiv.org/abs/hep-ph/0406278}
  {\path{arXiv:hep-ph/0406278}}, \href
  {http://dx.doi.org/10.1103/PhysRevC.71.064901}
  {\path{doi:10.1103/PhysRevC.71.064901}}.

\bibitem{Xu:2007aa}
Z.~Xu, C.~Greiner, {Transport rates and momentum isotropization of gluon matter
  in ultrarelativistic heavy-ion collisions}, Phys. Rev. C76 (2007) 024911.
\newblock \href {http://arxiv.org/abs/hep-ph/0703233}
  {\path{arXiv:hep-ph/0703233}}, \href
  {http://dx.doi.org/10.1103/PhysRevC.76.024911}
  {\path{doi:10.1103/PhysRevC.76.024911}}.

\bibitem{Uphoff:2010sh}
J.~Uphoff, O.~Fochler, Z.~Xu, C.~Greiner, {Heavy quark production at RHIC and
  LHC within a partonic transport model}, Phys. Rev. C82 (2010) 044906.
\newblock \href {http://arxiv.org/abs/1003.4200} {\path{arXiv:1003.4200}},
  \href {http://dx.doi.org/10.1103/PhysRevC.82.044906}
  {\path{doi:10.1103/PhysRevC.82.044906}}.

\bibitem{Sjostrand:2006za}
T.~Sjostrand, S.~Mrenna, P.~Skands, {PYTHIA 6.4 physics and manual}, JHEP 05
  (2006) 026.
\newblock \href {http://arxiv.org/abs/hep-ph/0603175}
  {\path{arXiv:hep-ph/0603175}}.

\bibitem{Uphoff:2010fz}
J.~Uphoff, O.~Fochler, Z.~Xu, C.~Greiner, {Production and elliptic flow of
  heavy quarks at RHIC and LHC within a partonic transport model}, J. Phys.
  Conf. Ser. 230 (2010) 012004.
\newblock \href {http://arxiv.org/abs/1004.4091} {\path{arXiv:1004.4091}},
  \href {http://dx.doi.org/10.1088/1742-6596/230/1/012004}
  {\path{doi:10.1088/1742-6596/230/1/012004}}.

\bibitem{Gossiaux:2008jv}
P.~B. Gossiaux, J.~Aichelin, {Towards an understanding of the RHIC single
  electron data}, Phys. Rev. C78 (2008) 014904.
\newblock \href {http://arxiv.org/abs/0802.2525} {\path{arXiv:0802.2525}},
  \href {http://dx.doi.org/10.1103/PhysRevC.78.014904}
  {\path{doi:10.1103/PhysRevC.78.014904}}.

\bibitem{Peshier:2008bg}
A.~Peshier, {Turning on the Charm}\href {http://arxiv.org/abs/0801.0595}
  {\path{arXiv:0801.0595}}.

\bibitem{Uphoff:2010sy}
J.~Uphoff, O.~Fochler, Z.~Xu, C.~Greiner, {Production, elliptic flow and energy
  loss of heavy quarks in the quark-gluon plasma}\href
  {http://arxiv.org/abs/1008.1995} {\path{arXiv:1008.1995}}.

\bibitem{Bouras:2010yw}
I.~Bouras, et~al., {Collective Flow and Energy Loss with parton transport}\href
  {http://arxiv.org/abs/1011.5073} {\path{arXiv:1011.5073}}.

\bibitem{Dokshitzer:1995qm}
Y.~L. Dokshitzer, G.~Marchesini, B.~R. Webber, {Dispersive Approach to
  Power-Behaved Contributions in QCD Hard Processes}, Nucl. Phys. B469 (1996)
  93--142.
\newblock \href {http://arxiv.org/abs/hep-ph/9512336}
  {\path{arXiv:hep-ph/9512336}}, \href
  {http://dx.doi.org/10.1016/0550-3213(96)00155-1}
  {\path{doi:10.1016/0550-3213(96)00155-1}}.

\bibitem{Peterson:1982ak}
C.~Peterson, D.~Schlatter, I.~Schmitt, P.~M. Zerwas, {Scaling Violations in
  Inclusive e+ e- Annihilation Spectra}, Phys. Rev. D27 (1983) 105.
\newblock \href {http://dx.doi.org/10.1103/PhysRevD.27.105}
  {\path{doi:10.1103/PhysRevD.27.105}}.

\bibitem{Xu:2008av}
Z.~Xu, C.~Greiner, {Elliptic flow of gluon matter in ultrarelativistic heavy-
  ion collisions}, Phys. Rev. C79 (2009) 014904.
\newblock \href {http://arxiv.org/abs/0811.2940} {\path{arXiv:0811.2940}},
  \href {http://dx.doi.org/10.1103/PhysRevC.79.014904}
  {\path{doi:10.1103/PhysRevC.79.014904}}.

\bibitem{Bouras:2008ip}
I.~Bouras, L.~Cheng, A.~El, O.~Fochler, J.~Uphoff, Z.~Xu, C.~Greiner, {Viscous
  Effects on Elliptic Flow and Shock Waves}\href
  {http://arxiv.org/abs/0811.4133} {\path{arXiv:0811.4133}}.

\bibitem{Xu:2010cq}
Z.~Xu, C.~Greiner, {Dependence of elliptic flow on number of parton degrees of
  freedom}\href {http://arxiv.org/abs/1001.2912} {\path{arXiv:1001.2912}}.

\bibitem{Fochler:2008ts}
O.~Fochler, Z.~Xu, C.~Greiner, {Towards a unified understanding of
  jet-quenching and elliptic flow within perturbative QCD parton transport},
  Phys. Rev. Lett. 102 (2009) 202301.
\newblock \href {http://arxiv.org/abs/0806.1169} {\path{arXiv:0806.1169}},
  \href {http://dx.doi.org/10.1103/PhysRevLett.102.202301}
  {\path{doi:10.1103/PhysRevLett.102.202301}}.

\bibitem{Fochler:2010wn}
O.~Fochler, Z.~Xu, C.~Greiner, {Energy loss in a partonic transport model
  including bremsstrahlung processes}\href {http://arxiv.org/abs/1003.4380}
  {\path{arXiv:1003.4380}}.

\end{thebibliography}

\end{document}